# Anisotropic Phonon Dynamics and Directional Transport in Actinide van der Waals Semiconductor USe$_3$

Aljoscha Söll[1*], Valentino Jadrisko[2*], Sourav Dey[3], Nassima Benchtaber[3], Kalyan Sarkar[1], Borna Radatovic[1], Jan Luxa[1], Fedor Lipilin[1], Kseniia Mosina[1], Vojtech Kundrat[4], Jakub Zalesak[5], Jana Vejpravova[6], Martin Zacek[6], Christoph Gadermaier[2], José J. Baldoví[3], Zdeněk Sofer[1§]

[1]*Department of Inorganic Chemistry, University of Chemistry and Technology Prague, Technicka 5, 166 28 Prague 6, Czech Republic*
[2]*Dipartimento di Fisica, Politecnico di Milano Piazza Leonardo da Vinci 32, 20133 Milano, Italy*
[3]*Instituto de Ciencia Molecular (ICMol), Universitat de Valencia, Paterna 46980, Spain*
[4]*Department of Chemistry, Faculty of Science, Masaryk University, Kamenice 5, Brno 62500, Czechia*
[5]*Chemistry and Physics of Materials, University of Salzburg, Jakob-Haringer-Strasse 2A, 5020, Salzburg, Austria*
[6]*Department of condensed matter physics, Faculty of mathematics and Physics, Charles University, Ke Karlovu 5, 121 16, Prague 2, Czech Republic*

*\* These authors contributed equally to this work.*
*§ Correspondence should be addressed to: Zdenek Sofer ([soferz@vscht.cz](soferz@vscht.cz))*

## ABSTRACT

Direction-dependent charge transport and optical responses are characteristic of van der Waals (vdW) materials with strong in-plane anisotropy. While transition-metal trichalcogenides (TMTCs) exemplify this behavior, heavier analogs remain largely unexplored. In this study we examine USe$_3$ as an anisotropic vdW material and a heavier analog of the well-studied TMTCs. We reveal strong in-plane anisotropy using polarization-resolved Raman spectroscopy, investigate strain-induced shifts of phonon modes, and quantify direction-dependent charge-carrier mobility through transport measurements on field-effect devices. First-principles calculations based on density-functional theory corroborate our findings, providing a theoretical basis for our experimental observations. Casting USe$_3$ as an actinide analog of a TMTC establishes a platform for exploring low-dimensional semiconductors that combine strong in-plane anisotropy with *f*-electron physics.

The emergence of graphene has ignited intense interest in layered and low-dimensional materials. Among these, anisotropic two-dimensional (2D) crystals are particularly attractive because their reduced symmetry leads to direction-dependent physical properties. Black phosphorus, rhenium dichalcogenides and the quasi-one-dimensional (1D) van der Waals (vdW) materials $TiS_3$ and $ZrTe_5$ all exhibit strong polarization-dependent optical responses and highly anisotropic electrical conduction.[1–6] Such behavior enables devices that can distinguish the polarization or propagation direction of light and charge carriers. For example, few-layer $TiS_3$ photodetectors show dichroic photoresponses, and black-phosphorus transistors achieve high on–off ratios by exploiting their in-plane anisotropy.[7–11]

Transition-metal trichalcogenides (TMTCs) provide a crystalline platform where quasi-1D physics can be explored in a 2D framework. These materials consist of infinite trigonal $MX_6$ chains (M = Ti, Zr, Hf, V, Nb, Ta; X = S, Se, Te) running along the crystallographic b-axis. Strong covalent bonding along the chains and comparatively weak bonding between chains yields van-der-Waals layers, which stack to form bulk crystals. This structural motif leads to unique combinations of properties: TMTCs exhibit charge-density waves, superconductivity and variable band gaps.[12–15] Their transport is extremely anisotropic; time-domain thermoreflectance measurements show that the thermal conductivity of $TiS_3$ along the chain axis is roughly twice that in the perpendicular direction.[16]

$ZrSe_3$ is a quasi-one-dimensional semiconductor with a near-direct band gap in the visible–NIR range and relatively high carrier mobilities, that has already been implemented in photodetectors and transistor-like device structures.[17–19] The robust anisotropy of TMTCs can be tuned by external stimuli such as pressure, temperature and tensile strain,[20] opening opportunities for thermoelectric applications and mechanically responsive electronics. These specific examples demonstrate that quasi-1D materials offer more than just directional curiosities; they are an emerging class of semiconductors with tunable, technologically relevant properties.[21] Extending this chemistry beyond transition metals introduces heavier actinide analogs. Recent work on uranium-based 5*f* compounds already reveals exotic quantum phases. In the heavy-fermion metal UTe$_2$, scanning tunnelling microscopy detects chiral in-gap states at step edges, consistent with a spin-triplet topological superconducting state and providing a promising platform for Majorana edge modes.[22] By contrast, USe$_3$ remains essentially unstudied beyond a few preliminary observations. It adopts the same monoclinic structure as ZrSe$_3$, comprising edge-sharing [USe$_3$]$_\infty$ chains arranged into van der Waals layers. Early magnetic and transport measurements reported that USe$_3$ behaves as an insulator,[23] and density-functional calculations incorporating on-site Coulomb interactions predict an antiferromagnetic ground state with an indirect band gap.[24] Yet little is known about its optical or electronic anisotropy, and its mechanical response to strain remains unexplored.

In this work, following a recent push to use dimensional control and local strain to access quantum behavior in actinide materials,[25] we revisit USe$_3$ from the perspective of quasi-1D materials science. We combine density-functional theory with experiments on exfoliated flakes to probe its electronic structure, vibrational anisotropy, strain-tunable phonons, and charge transport. By casting USe$_3$ as an actinide analog of well-studied TMTCs, we open a pathway for understanding and exploiting the low-dimensional properties of heavy-element chalcogenides and provide a solid framework for future exploration of actinide-based low-dimensional semiconductors.

**RESULTS AND DISCUSSION**

USe$_3$ crystallizes in the P2$_1$/m space group, where [USe$_3$]$_\infty$ chains run along the *b*-direction and stack through van der Waals (vdW) interactions into layered sheets (Figure 1 a-c). This structural anisotropy points to strongly direction-dependent electronic and phononic responses. We therefore begin by integrating density-functional theory with bulk diffraction and spectroscopy to establish the crystal and electronic structure. Next, we fabricate and characterize 2D USe$_3$-based devices to quantify anisotropic phonon shifts under uniaxial strain, and finally we probe charge transport along the two in-plane axes.

**Crystal and Electronic Structure**

We calculated the electronic band structure of bulk USe$_3$ using spin polarized DFT + U (See Methods). Figure 1d reveals an indirect 1.07 eV band gap with the highest valence bands and lowest conduction bands of USe$_3$ distributed along B−Γ−A and A−E, respectively. Projected density of states (PDOS) indicated that both Se and U electronic states contribute to the valence band maximum (VBM), while the conduction band minimum (CBM) is mainly dominated by uranium electronic states (Figure 1e). We did not consider spin-orbit coupling (SOC) in our calculations as its effect on the electronic structure was found to be minimal in the previous studies.[24] Uniaxial strain of <1% leads to a decrease in band gap, while the same strain along *b* increases it (Supplementary Fig. S1), giving a first indication that structural anisotropy exerts a direct influence on the electronic properties. Finally, Figure 1f shows the phononic band structure of USe$_3$.

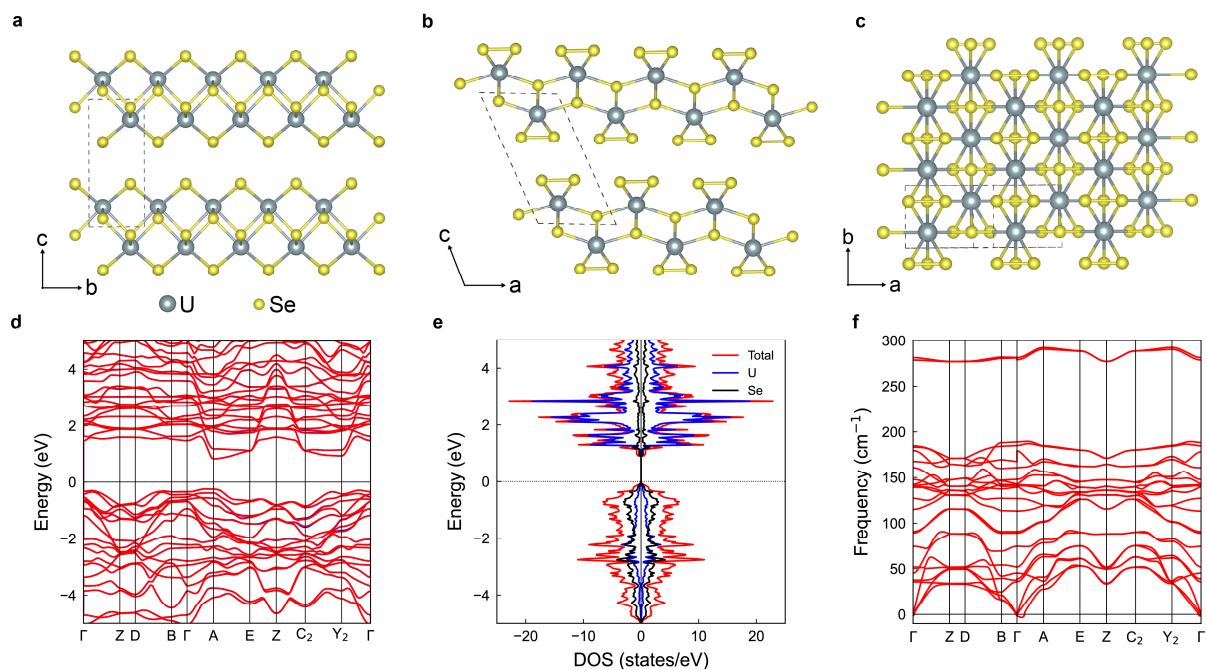

**Figure 1**. Crystal structure and computed electronic structure of USe$_3$. (a-c) Crystal structure viewed along *a, b,* and *c*-directions.[26] (d) Electronic band structure of USe$_3$. (e) Projected Density of states (DOS). (f) Phonon band structure.

Scanning electron microscopy (SEM) images of a single crystal of USe$_3$ are shown in Figure 2a. The magnified sections in 2b shows a terraced morphology on the surface showcasing the layered character of the material. Towards the bottom end of the crystal parts of the surface layers have delaminated and are pointing upwards, an artifact from cleaving the surface immediately prior to the measurement. Here we observe that the cleaved layers have a preferred orientation and break more easily along the *a*-directions, a direct result of the structural anisotropy.

Energy dispersive X-ray spectroscopy (EDS) maps shown in Figure 2c reveal a uniform distribution of uranium and selenium throughout the crystal with a slight excess of uranium and negligible amounts of oxygen on the surface. The elemental composition across the crystal was U$_{1.16}$Se$_3$ (Supporting Information, Section 3).

We measured X-ray diffraction on a USe$_3$ single crystal instead of powder due to the air and moisture sensitivity of the material. The resulting spectrum is shown in Figure 2d shows excellent crystallinity of the sample. As the crystal was lying flat, only peaks corresponding to the 00L indices are visible which correspond to XRD measurements of USe$_3$ as reported in the literature (Supporting Information, Section 2). There is one additional peak marked by * which we ascribe to elemental selenium forming on the surface of USe$_3$ via photo-enhanced oxidation(Supporting Information, Section 2).[27]

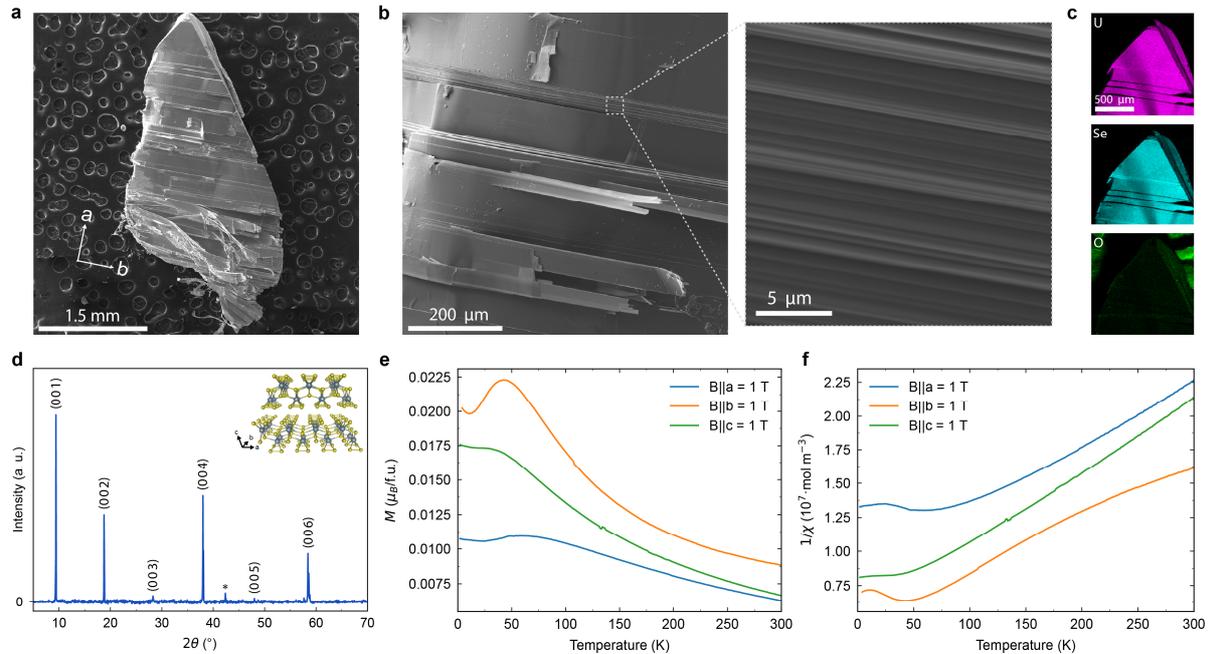

**Figure 2**. (a) Scanning electron microscopy (SEM) micrograph of USe$_3$ single crystal with a lateral dimension of ~5 mm. The in-plane lattice axes are marked in the image, a preferred orientation for crystal edges and delamination is visible, with sheets elongated in the *b*-direction. (b) Magnified SEM micrographs reveal terraced morphology indicative of a layered structure. (c) Energy dispersive X-ray spectroscopy (EDS) maps for the elements U, Se, and O taken on the top part of the crystal in a. (d) X-ray diffraction pattern of flat USe$_3$ crystal, strong 00L peaks indicate good crystallinity. One unassigned peak is marked by * likely originating from β-cubic Se (e) Temperature-dependence of magnetization $M(T)$ of USe$_3$ single-crystal for $B = 1\,\text{T}$ applied along the three principal crystallographic axes. (f) Temperature-dependence of the inverse molar magnetic susceptibility $1/\chi(T)$.

Detailed magnetization measurements were performed along the principal crystallographic directions to capture the anisotropy in magnetic susceptibility and key magnetic parameters. Figure 2e shows the temperature dependence of magnetization $M(T)$ measured at the applied magnetic field of $B = 1$T. All curves exhibit a broad maximum consistent with the onset of long-range antiferromagnetic ordering, within the range reported in the literature. Importantly, the peak temperature shifts by ~10 K between orientations, demonstrating the material's magnetic anisotropy. The response magnitude is also orientation-dependent, with $M$ largest for $B||b$ suggesting that $b$ is the easy axis at this field. Figure 2f shows the molar magnetic susceptibility in inverse form, where a clear deviation from the Curie–Weiss behavior is observed, likely reflecting strong crystal-field contributions.[28–31] Further discussion is provided in Supporting Information Section 4.

The USe$_3$ crystal was analyzed by transmission electron microscopy (TEM) and selected area electron diffraction (SAED) techniques. The chipped USe$_3$ crystal was initially inspected along the <001> axis, revealing an ordered structure (Figure 3a). The SAED analysis (Figure 3b) of the same spot on the crystal provided an expected diffraction pattern closely matching the simulated analysis (Single Crystal software). The diffraction in Figure 3b also contains a minor secondary pattern of USe$_3$ (marked by yellow circles) that is produced by differently angled crystallites on the edge of the analyzed area (not shown in Figure 3a). The bulk crystal was analyzed by SEM (Figure 3c). Similarly to the SEM study in Figure 2 a-b, the crystal sample observed shows an apparent layered structure with tendencies to break apart into strap or band-like shards. Subsequently, the USe$_3$ crystal was inspected as a lamella lifted out from the bulk sample, with its position marked in the SEM image Figure 3c by a yellow line. The orientation of the focused ion beam (FIB) cut was selected based on the polarized light Raman spectroscopy. The TEM analysis of the cross-section (Figure 3d) shows the layered structure of the USe$_3$ with the indicated interlayer distance of 1.02 nm. The closer analysis of the lamella shows distinct surface oxidation that correlates with the air instability and photo-enhanced oxidation of the material. The TEM analysis of the oxidized interface is shown in Figure S4. The pristine USe$_3$ phase in the cross-section was inspected by SAED analysis and compared with the computed patterns (Figure 3e). The lamella was angled along various orientations to prove the structural order in the crystal, while using a double tilt TEM holder. The sequence of angle-resolved SAED analyses precisely matches the calculated diffraction patterns.

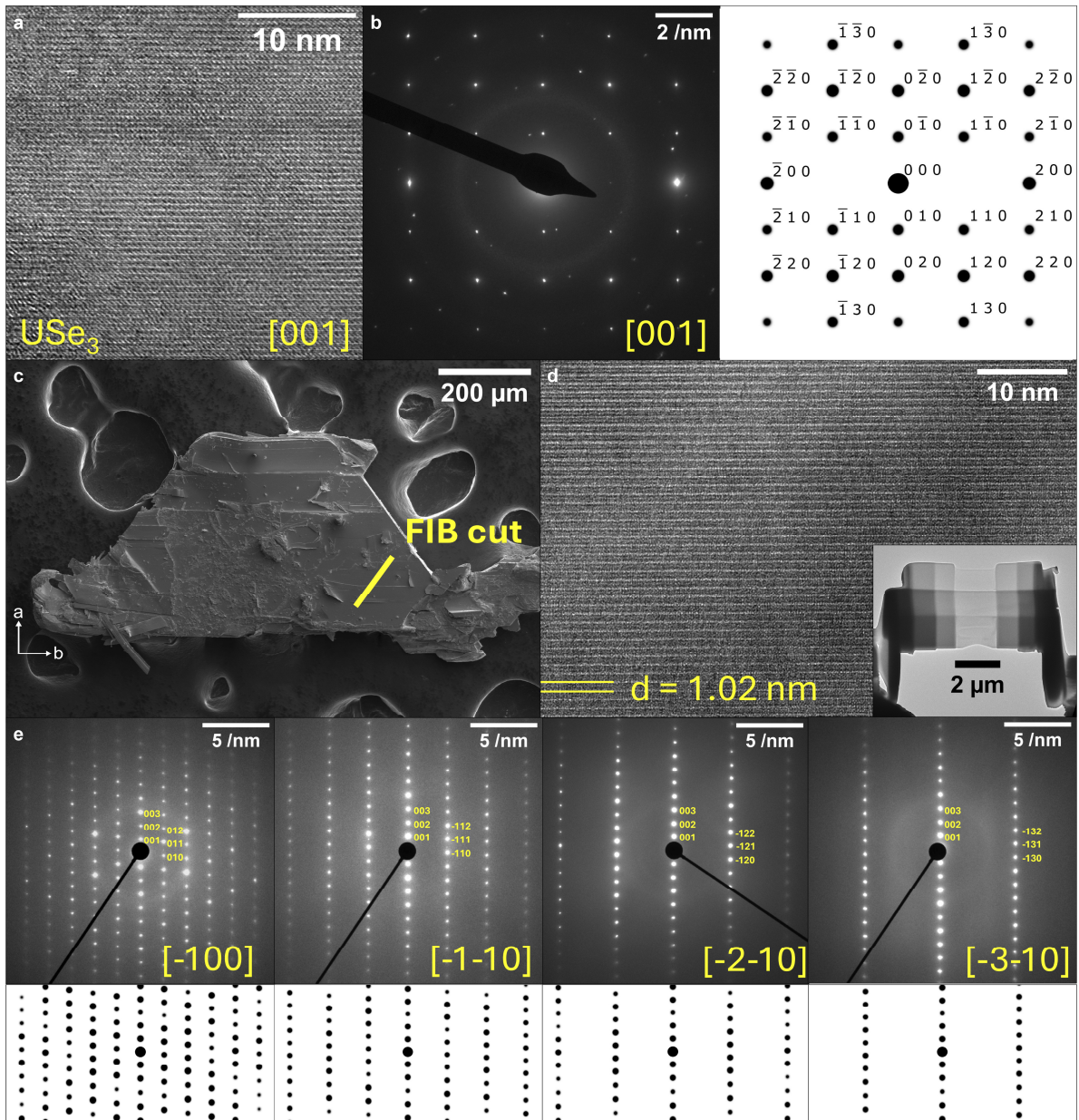

**Figure 3**. Structural analysis of the synthesized USe$_3$ crystal. (a) TEM analysis of the thin USe$_3$ crystal along the [001] vector. (b) the corresponding SAED analysis of the spot in (a) closely fits the calculated SAED pattern shown on the right panel. (c) SEM analysis of the USe$_3$ crystal with a yellow line-marked area lifted out as a lamella. (d) TEM analysis of the lamella shown in the inset and its position marked in (c). The interlayer distance was measured as 1.02 nm. (e) The series of SAED measurements performed on the lamella from (d), showing different orientations obtained by tilting of the lamella. The SAED analyses are compared with matching simulated SAED patterns shown in the line below.

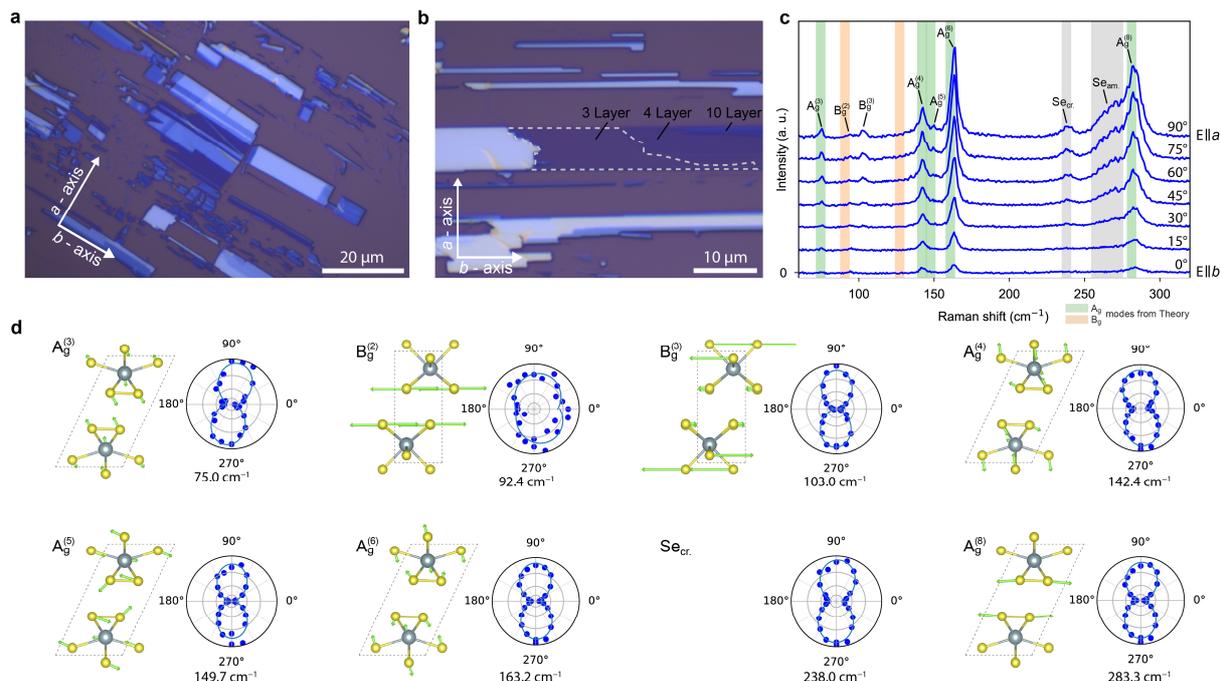

**Figure 4**. **a**. Exfoliation and polarization-resolved Raman of USe$_3$ (a) Optical Micrograph of exfoliated USe$_3$ on Si/SiO$_2$ showing aligned rectangular sheets with a preferred orientation along the *b*-direction. (b) Magnified image shows few-layer regions (≈3–4 L) tens of micrometers across. (c) Raman spectra at room temperature for rotations of the incident linear polarization in the *ab*-plane (0°: E∥*b*; 90°: E∥*a*); green/orange bands mark calculated A$_g$/B$_g$ modes, gray bands indicate crystalline and amorphous Se. Seven of twelve Raman-active modes are observed. (d) Mode schematics and polarization polar plots (0°: E∥*b*; 90°: E∥*a*).

**Vibrational anisotropy and Strain-Dependence**

Our vibrational characterization was done on exfoliated sheets prepared in an argon atmosphere. USe$_3$ exfoliates well, producing rectangular sheets elongated in the *b*-direction. The behavior and yield are comparable to other anisotropic materials such as CrSBr[32] or ZrSe$_3$. Figure 4a and b show exfoliated sheets on Si/SiO$_2$, containing few-layer sheets as thin as 3 nm (~3 Layers) with lateral dimensions in the order of tens of micrometers.

We measured Raman on the exfoliated samples and assigned the observed Raman modes in accordance with our computational theory. According to group theory, there are 12 Raman-active vibrational modes in USe$_3$: 8 A$_g$ (vibrations inside the mirror plane, inside the *ac*-plane) and 4 B$_g$ modes (vibrations outside the mirror plane, along the *b*-direction). We number the modes A$_g^{(1)}$ to A$_g^{(8)}$ and B$_g^{(1)}$ to B$_g^{(4)}$ and provide a list of computed modes in table S1, here all modes are positive, confirming the dynamical stability of USe$_3$. From the 12 predicted Raman modes we observe seven at room temperature: 2 B$_g$ modes at B$_g^{(2)}$: 94.2 cm$^{-1}$ and B$_g^{(3)}$: 103.0 cm$^{-1}$ and 5 A$_g$ modes at A$_g^{(3)}$: 75.0 cm$^{-1}$, A$_g^{(4)}$: 142.4 cm$^{-1}$, A$_g^{(5)}$: 149.7 cm$^{-1}$, A$_g^{(6)}$: 163.2 cm$^{-1}$ and A$_g^{(8)}$: 283.4 cm$^{-1}$. There is one additional mode at 239.1 cm$^{-1}$ and a broad shoulder around 270 cm$^{-1}$ which are identified in the literature as originating from crystalline and amorphous selenium respectively, likely caused by photo-enhanced surface oxidation of the sample.[27,33] Our theoretical predictions are overall in good agreement with the experiment (Figure 4c, Supplementary section 5). We do not observe any significant thickness dependence of the Raman peak positions between few-layer and bulk samples (Supporting Information, Section 6).

When rotating the incident linear polarization in the *ab*-plane we observe pronounced optical anisotropy as seen in Figure 4c. Polar plots in Figure 4d reveal a clear twofold symmetry for every vibration except B$_g^{(2)}$; all other modes reach maximum intensity when the electric field is aligned with the *a*-axis and are nearly extinguished for E∥b. Fully polarization-resolved Raman measurements further show that intensity is strongest in parallel geometries (Z(XX)$\overline{Z}$, Z(YY)$\overline{Z}$)) and suppressed in crossed geometries (Z(XY)$\overline{Z}$, Z(YX)$\overline{Z}$) (Figure S8b). This observation is fully consistent with our DFPT-calculated Raman tensors, which are dominated by the first element along the diagonal (Supporting Information, Section 7). Such behavior is typical for highly anisotropic and quasi-1*D* materials where the Raman tensor is often dominated by one of the diagonal elements. In contrast, the B$_g^{(2)}$ mode exhibits significant off-diagonal tensor elements in our calculations, and the polarization-resolved measurements in Figure S8b confirm that its intensity is essentially independent of the incident-polarization angle. Unlike the other modes, it is also most intense in crossed geometries and suppressed in parallel geometries, consistent with off-diagonal Raman-tensor elements

Finally, the modes corresponding to elemental selenium show the same two-fold symmetry as the USe$_3$ itself, indicating they originate from orientation-preserving surface oxidation of USe$_3$, rather than originating from isotropic elemental Se residues left over from the synthesis.

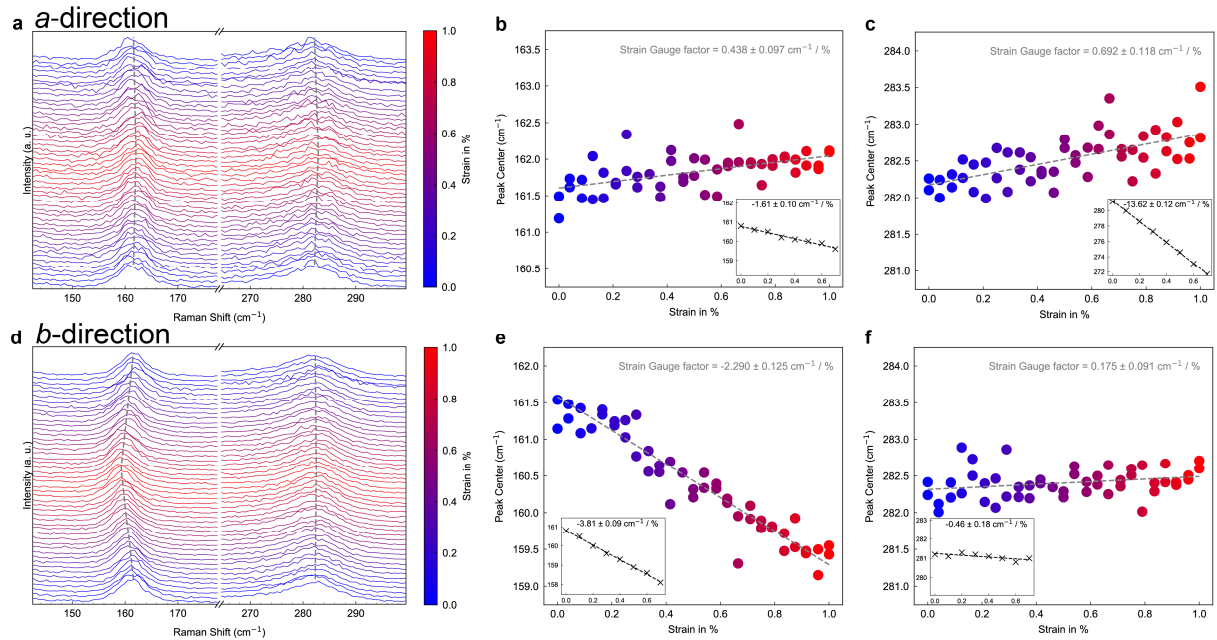

**Figure 5.** Raman shifts under uniaxial tensile strain. (a) Waterfall plot of $A_g^{(6)}$ and $A_g^{(8)}$ for strain along *a* (0 to 1%; color encodes strain, from bottom to top). (b,c) Peak position vs. strain in *a* with linear fits. Inset shows theoretical prediction. (d) Waterfall plot for strain along *b*. (e,f) Peak position vs. strain in *b* with linear fits. Inset shows theoretical prediction. Measurements on exfoliated flakes on polycarbonate substrates at room temperature in Ar.

We induced uniaxial tensile strain up to 1% in exfoliated sheets of USe$_3$ using a three-point bending setup inside an argon filled glovebox and measured the effect using Raman spectroscopy. Figure 5 shows the effect of tensile strain on the two most intense A$_g$ modes ($A_g^{(6)}$ and $A_g^{(8)}$) along the *a* and *b* crystallographic axes. The respective sections of the Raman spectra are plotted in Figure 5a and d for different values of uniaxial strain with the extracted peak center position directly plotted against the strain value in comparison to the theoretical prediction in Figure 5b-c and e-f for the *a* and *b*-direction respectively.

For tensile strain along the *b*-direction we can clearly observe a red shift of the out-of-plane $A_g^{(6)}$ mode, by ca. 2.2 cm$^{-1}$ at the maximum strain of 1% which corresponds to a strain gauge factor of $-2.29 \pm 0.13$ cm$^{-1}$/%. In contrast, the mostly in-plane $A_g^{(8)}$ mode shows a negligible shift, the peak position stays nearly constant with a strain gauge factor of $0.18 \pm 0.11$ cm$^{-1}$/%. Both results are in good accordance with our theoretical calculations which predict a redshift for $A_g^{(6)}$ but no significant shift for $A_g^{(8)}$.

However, when we apply uniaxial strain along the *a*-axis to the same sample, the two $A_g^{(6)}$ and $A_g^{(8)}$ modes shift weakly towards higher energies: $0.44 \pm 0.10$ cm$^{-1}$/% and $0.69 \pm 0.12$ cm$^{-1}$/%, respectively. This small blue shift contrasts with first principle calculations, which predict a modest red shift for the first and a significant red shift for the second vibration (all values are compared in Table 1, additional measurements in Supporting Information, Section 8).

Optical microscopy of the same flake under ~1% applied strain (Figure S11) reveals that tensile strain along *a* is not transmitted uniformly into the lattice. Instead, the crystal relaxes by forming multiple parallel cracks and delaminates into narrow ribbons. This fracture-mediated strain relief reflects the anisotropic nature of the crystal lattice with weaker interatomic bonding along *a* compared to the *b*-direction. The sample mechanically fails before significant strain can develop within the lattice, thus suppressing the phonon frequency shift that theory predicts for a uniformly strained lattice. The slight blue shift which is observed in our experiment is likely a result of initial non-uniform strain within the structure from the fabrication, which is released when the material delaminates. This causes the already slightly strained structure to relax leading to a slight peak narrowing and blue shifting of Raman modes in some of the measured samples (Supporting Information, Section 9).

Table 1. Comparison between experimental and theoretical strain gauge factors in the two principal directions.

|  | *a* | | *b* | |
|---|---|---|---|---|
|  | $A_g^{(6)}$: 163.2 cm$^{-1}$ | $A_g^{(8)}$: 283.3 cm$^{-1}$ | $A_g^{(6)}$: 163.2 cm$^{-1}$ | $A_g^{(8)}$: 283.3 cm$^{-1}$ |
| Experimental (cm$^{-1}$/%) | $0.44 \pm 0.10$ | $0.69 \pm 0.12$ | $-2.29 \pm 0.13$ | $0.18 \pm 0.11$ |
| Computed (cm$^{-1}$/%) | $-1.61 \pm 0.10$ | $-13.62 \pm 0.12$ | $-3.81 \pm 0.09$ | $-0.46 \pm 0.18$ |

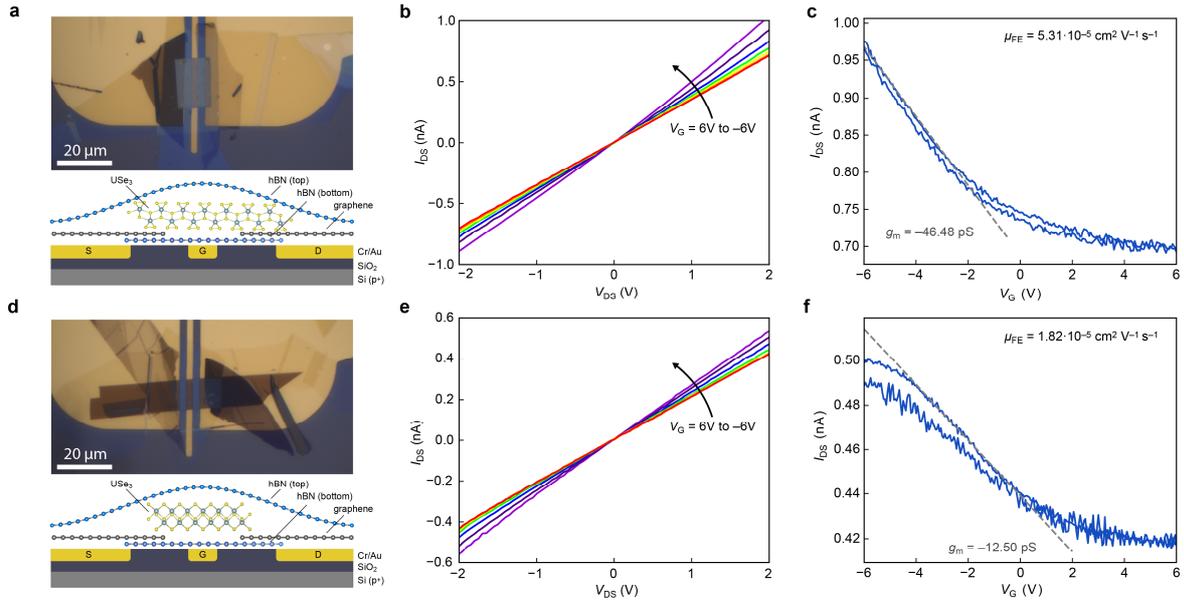

**Figure 6.** Charge transport in hBN-encapsulated USe$_3$ FETs with graphene contacts. (a) Optical micrograph and schematic of FET with a channel oriented along *a*. (b) Output characteristics $I_{DS}$ vs. $V_{DS}$ for $V_G$ swept from +6 V to −6 V. (c) Transfer characteristics $I_{DS}$ vs. $V_G$ at $V_{DS}$ = 2 V showing p-type carrier modulation. (d) Optical micrograph and schematic along *b*. (e) Output characteristics $I_{DS}$ vs. $V_{DS}$ for $V_G$ = +6 V to −6 V. (f) Transfer characteristics $I_{DS}$ vs. $V_G$ at $V_{DS}$ = 2 V showing p-type carrier modulation.

**Directional Charge Transport**

According to our band structure calculations, USe$_3$ is a semiconductor with a bandgap in the near-infrared (NIR) range, making it a potential channel material for field-effect transistors. To test its performance as a channel material along the *a*- and *b*-directions, we fabricated two hBN-encapsulated field-effect transistors (FETs) using USe$_3$ as a channel in two different orientations. To improve contact resistance and avoid the formation of Schottky barriers we use few-layer graphene as source and drain contacts. Figures 6a and d show optical micrographs of the devices in the *a*- and *b*-directions, respectively as well as a schematic of the heterostructure. For both devices, we observe very linear $I_{DS}$ vs. $V_{DS}$ behavior in the output curve shown in Figure 6b and e for the *a*- and *b*-direction respectively. Highly linear output characteristics indicate ohmic contacts between the USe$_3$ and few-layer graphene, which is often observed with graphene as a contact material for 2D semiconductors, due to graphene's ability to form atomically sharp and clean vdW interfaces and favorable work function alignment. These unique properties reduce Fermi-level pinning and electrostatic work function modulation at the contact, leading to low-resistance and ohmic contacts.[34,35] However, the overall measured current is low, on the order of up to one nanoampere in either direction. When a gate voltage is applied, we observe modulation of the source-drain current, indicating p-type conductivity. To quantify this effect, we measured the transport characteristics as shown in Figure 6c for the *a*-direction and Figure 6f for the *b*-direction and we derive the transconductance $g_m$ in the linear regime by calculating the slope of the linear section according to $g_m = \frac{\partial I_{DS}}{\partial V_{GS}}$. From the transconductance of $g_{m,a} = -46.48 \text{ pS}$ and $g_{m,b} = -12.50 \text{ pS}$ extracted at a small $V_{DS}$ in the linear regime, we calculate a field-effect mobility according to $\mu_{FE} = \frac{L_{ch}|g_m|}{W_{ch} C_{ins} V_{DS}}$ of $\mu_{FE,a} = 5.31 \cdot 10^{-5} \text{ cm}^2 \text{ V}^{-1} \text{ s}^{-1}$ and of $\mu_{FE,b} = 1.82 \cdot 10^{-5} \text{ cm}^2 \text{ V}^{-1} \text{ s}^{-1}$ for the *a*- and *b*-directions, respectively. Here $L_{ch}$ is the channel length, $W_{ch}$ the channel width, $C_{ins}$ the insulator capacitance of hBN, and $V_{DS}$ the source-drain voltage during the measurement. For clarity these values are specified in Table S4. The anisotropy ratio of mobility for the two respective directions is around $\frac{\mu_{FE,a}}{\mu_{FE,b}} \approx 2.9$ indicating an anisotropy of mobility for USe$_3$ with a higher mobility in the *a*-direction.

To get a theoretical foundation for the transport in USe$_3$ we modeled the transport behavior computationally. The transport computation setup is shown in Figure S14. The source and drain of the system are semi-infinite metallic leads that are coupled to a central USe$_3$ area. The carrier concentration is modulated via a gate electrode. Electrical conductivity exhibits substantial anisotropic behavior when the transport computations are carried out along the *a* and *b* crystallographic directions. In both cases p-type transport is observed and the computed mobility values are $\mu_a = 4.85 \cdot 10^{-4} \text{ cm}^2 \text{ V}^{-1} \text{ s}^{-1}$ and $\mu_b = 1.62 \cdot 10^{-4} \text{ cm}^2 \text{ V}^{-1} \text{ s}^{-1}$ indicating that transport is significantly more efficient along the *a* direction than the *b* direction with a computed anisotropy ratio of the mobility of $\frac{\mu_a}{\mu_b} \approx 3.0$ which is very close to the experimental ratio of $\frac{\mu_{\text{FE},a}}{\mu_{\text{FE},b}} \approx 2.9$.

Overall computational theory and experiment are in good agreement here, both indicating a higher mobility in the *a*-direction than the *b*-direction by a similar factor. For both directions the absolute value of the mobility is within the same order of magnitude with the computational value being slightly higher than the experimental one. This is to be expected, the computational value is closer to the intrinsic mobility of the material while the two-terminal FET which was fabricated, only allows us to measure the field-effect mobility $\mu_{\text{FE}}$, which is typically lower than the intrinsic mobility since detrimental effects such as the contact resistance are not taken into account.[36]

In both cases, however, the carrier mobility in either direction is still several orders of magnitude lower than in other 2D semiconductors such as $MoS_2$.[36] To explain why, we performed effective mass analysis of the valence bands from our DFT band structures. Using a quadratic fit around the valence band maxima, we obtained large hole effective masses of $m_a^* \approx 1.841\ m_0$ along the *a*-direction and $m_b^* \approx 2.072\ m_0$ along the *b*-direction (Supporting Information, effective mass determination). The results indicate slightly lighter effective masses along the a-direction, in line with the experimental mobility trend. At the same time, both values are still several times larger than those of prototypical 2D semiconductors such as $MoS_2$ with (effective hole mass of $m^* \approx 0.5\ m_0$[37,38]), supporting the observed low mobilities, according to $\mu = e\tau/m^*$ (here *e* is the elementary charge, $\tau$ the scattering time, and $m^*$ the effective hole mass). In addition, the low density of states at the band edges, visible in the calculated DOS (Figure 1e), reduces the carrier concentration for a given gate voltage. In regimes where charged-impurity scattering dominates, a low carrier density further weakens Coulomb screening, shortening the scattering time $\tau$ and thereby decreasing $\mu$.[39] Both effects can be traced back to the flat, weakly dispersive uranium-derived bands at the Fermi level, which lead to strong carrier localization, as also discussed in the broader context of f-electron materials.[40] Thus, the reduced mobility in USe$_3$ is not simply a device artifact (e.g. contact resistance) but an intrinsic consequence of its band structure, and our DFT-based effective mass analysis quantitatively supports the experimental observations.

**CONCLUSION**

USe$_3$ emerges as a van der Waals material with strong in-plane anisotropy and as an actinide analog of a transition metal trichalcogenide. Polarization-resolved Raman spectroscopy reveals strong in-plane anisotropy of phonon modes governed by diagonal tensor elements. Controlled uniaxial tensile strain affects phonon frequencies, with a clear red shift of $A_g^{(6)}$ for strain along *b*, with a strain gauge factor of −2.29 cm$^{-1}$/% comparable to the upper range of values reported for benchmark systems such as monolayer MoS$_2$ under uniaxial strain,[41] but a muted response along *a* due to crack-mediated relaxation. Graphene-contacted field-effect devices show p-type conduction and a mobility anisotropy $\mu_a/\mu_b$ near three. First-principles calculations reproduce the vibrational and transport trends and attribute the low absolute mobilities to heavy U-derived bands with weak dispersion and a reduced density of states near the band edge. These results link crystal anisotropy to direction-dependent phonon dynamics and charge transport in a heavy-element trichalcogenides paving the way for integrating *f*-electron physics into anisotropic 2D device architectures.

**METHODS**

The First principles calculations were performed within the framework of density functional theory (DFT) as implemented in the Vienna ab initio simulation package (VASP).[42] The exchange-correlation was described with Perdew-Burke-Ernzerhof's (PBE) functional the of the generalized gradient approximation (GGA).[43] The projected augmented wave (PAW) method was employed in our calculations with the frozen core approximation. Van der Waals interactions were taken into account by means of the DFT-D3 method of Grimme. A Γ-centered 9×13×3 Monkhorst-Pack k-point mesh and plane wave cutoff energy of 400 eV was employed for structural optimisation converged to within $10^{-6}$ eV in total energy $10^{-3}$ eV/atom in total forces. A strong on-site Coulomb repulsion of 4.0 eV was employed using the formalism of Dudarev et al. to describe the localized 5f electrons of uranium.[24,44] Vibrational frequencies were estimated using the Phonopy code[45] using density functional perturbation theory (DFPT) with a 2×2×1 supercell. The IR and Raman active frequencies were obtained by analyzing the symmetry of the corresponding vibrational modes. To reduce the computational cost, the k-point mesh was set to $5 \times 7 \times 5$. The third-order force constant was estimated with a finite displacement approach from 80 independent atomic displacements as implemented in the ShengBTE code.[46,47] We have constructed a derived tight-binding model based on maximally localized Wannier functions as implemented in the Wannier90 code.[48] In this model, the reduced basis set was formed by 5f and 6d orbitals of uranium and 2p orbitals of Se. The magnetic exchange was estimated using Green's function approach as implemented in the TB2J code.[49] We use the NEGF formalism based on Density Functional Theory (DFT) calculations to explain quantum transport in $USe_3$ and Landauer formula.[50,51] The system's Hamiltonian is generated from the Wannier tight-binding Hamiltonian. The system's Green's function is calculated as follows: $G^r(E) = [(E+i\eta)I - H_{central} - \Sigma_L^r - \Sigma_R^r]^{-1}$ where $H_{central}$ is the Hamiltonian of the central region obtained from Wannier90. $\Sigma_L$ and $\Sigma_R$, are the self-energy corrections from the left and right leads, respectively, describing the influence of semi-infinite contacts. I is the identity matrix, and η is an infinitesimally small broadening term.

**Characterization**

Raman spectral measurements and mapping were conducted with a WITec Confocal Raman Microscope (WITec alpha300 R, Ulm, Germany), equipped with a 532 nm laser and a spectrometer with a thermoelectrically cooled CCD camera sensor. The measurements were performed in an argon-filled glovebox at room temperature with a 100× objective and a laser power of less than 2.5 mW to avoid sample degradation. To apply uniaxial strain Polycarbonate (PC) containing the sample substrate was bent using a motorized setup. Characterization by Atomic Force Microscopy (AFM) was performed on a NanoMagnetics ezAFM compact atomic force microscopy system in dynamic mode (tapping mode) inside an argon-filled glovebox. X-ray diffraction was carried out on a Bruker D8 Discover with Cu X-ray source ($\lambda = 0.15418$ nm, $U = 40$ kV, $I = 40$ mA). Diffractograms were collected in a range from 10° to 90° with a step of 0.02° and integration time of 0.2 s. The data was processed in HighScore plus software package. Diffractograms were then normalized to the most intense peak. The morphology of the samples was investigated using scanning electron microscopy (SEM) with a FEG electron source (Tescan Lyra dual beam microscope). The samples were placed on carbon conductive tape. SEM measurements were carried out by using a 5 kV electron beam. The composition of the samples was determined using an energy dispersive spectroscopy (EDS) analyzer (X-MaxN) with a 20 mm$^2$ SDD detector (Oxford Instruments). Data was evaluated using AZtecEnergy software. EDS measurements were carried out with a 15 kV acceleration voltage. Magnetization measurements were performed using the vibrating sample magnetometer (VSM) option integrated into a 14 T Physical Property Measurement System (Quantum Design, USA). Samples were placed in a low-background sample holder inside a glove box and promptly loaded into the measurement chamber to prevent degradation. Temperature and magnetic field dependencies were recorded with the magnetic field aligned along each of the principal axes of the monoclinic crystal structure. The temperature range and sweep rate was set to 2 – 300 K and 2 K/min, and the magnetic field range and sweep rate to 0 – 14 T and 5 mT/s, respectively. The measured values were converted to molar units (molar mass, Mr = 474.91 g/mol; number of formula units per unit cell, Z = 1) for further analysis. For SEM imaging (prior to TEM measurements) and FIB sample preparation, a Thermo Fisher Scientific Helios 5 FX was used. The USe3 crystal was transferred onto carbon tape and oriented according to the polarized light Raman spectroscopic data. Expected plane normal was [001]. Subsequently, the sample was rotated by 35°, and a cross-sectional FIB lamella was lifted out in a standard way. A Thermo Fisher Scientific Titan was utilized for the TEM and SAED analyses. The microscope was run at 300 kV in TEM mode. Room temperature transport measurements were carried out in ambient conditions with a two channel Keysight B2902A Precision Source/Measure Unit (SMU) in a 4-probe measurement setup.

**Sample preparation**

All exfoliated samples in this work were prepared using the scotch tape method. For Polarization resolved Raman spectroscopy the samples were directly exfoliated on $Si/SiO_2$ substrates. For strain measurements the samples were directly exfoliated on PC sheets with a thickness of 0.25 mm. For the device fabrication all materials were first exfoliated on PDMS (gelpack) to increase the yield and then picked up from PDMS using a PC membrane on a PDMS stamp. Specifically, hBN was picked up first and then used to pick up the remaining sheets. Subsequently, the stack was released on $Si/SiO_2$ (320 nm) substrates with prefabricated gold source drain and gate electrodes (buried contacts 5/100 nm Cr/Au), by increasing the substrate temperature to 180 °C. The PC membranes were cleaned by chloroform. Finally, all devices were annealed in high vacuum ($10^{-7}$ mbar) at 300 °C for 4 hours.

**SAFETY**

During this work all samples were handled inside an argon filled glovebox to avoid degradation of the moisture sensitive USe$_3$. U$^{238}$, the predominant uranium isotope, is an alpha emitter, which is easily blocked by the thick latex gloves of the glovebox. Furthermore, the amount of material which was typically used during exfoliation and sample preparation was in the order of a few milligrams. In such small quantities the risk of accidental irradiation or poisoning from the samples is negligible.

**AUTHOR CONTRIBUTIONS**

Z.S. initiated and supervised the project. A.S. and Z.S. grew the USe$_3$ crystals. A.S. and V.J. fabricated devices and conducted strain and Raman analysis with input from C.G.. A.S. performed transport measurements with input from K.S. and J.L.. S.D., N.B., and J.J.B. performed the computational calculations and analyzed the results. B.R. and F.L. performed SEM and EDS measurements and analyzed the data. K.S. prepared samples for the TEM measurements. V.K. and J.Z. performed TEM measurements and analyzed the data. A.S. prepared samples for low temperature magnetic susceptibility measurements, J.V. and M.Z conducted the measurement and analyzed the data. A.S. and V.J. wrote the manuscript, with input from all authors.

**COMPETING FINANCIAL INTERESTS**

The authors declare no competing financial interests.

## DATA AVAILABILITY

The data that support the findings of this study are available from the corresponding author on reasonable request.

## ACKNOWLEDGEMENTS

This work received funding from the European Union's Horizon 2020 research and innovation program under grant agreement 956813 (2Exciting). This project was also supported by the Czech Science Foundation Grant No. 24-11465S. JJB and SD would like to thank EU (Grant No. 2D-SMARTiES ERC-StG-101042680 and Marie Curie Fellowship SpinPhononHyb2D 101107713) and the Generalitat Valenciana (grant CIDEXG/2023/1) for the funding. The work was supported by a grant from the Programme JohannesAmos Comenius under the Ministry of Education, Youth and Sports of the Czech Republic CZ.02.01.01/00/22_008/0004558 Advanced MUltiscaLe materials for key Enabling Technologies. Magnetic measurements were performed in MGML (mgml.eu), which is supported within the program of Czech Research Infrastructures (project no. LM2023065).

# "Anisotropic Phonon Dynamics and Directional Transport in Actinide van der Waals Semiconductor USe3" Supporting Information

Manuscript Title: Anisotropic Phonon Dynamics and Directional Transport in Actinide van der Waals Semiconductor USe$_3$

Aljoscha Söll[1*], Valentino Jadrisko[2*], Sourav Dey[3], Nassima Benchtaber[3], Kalyan Sarkar[1], Borna Radatovic[1], Jan Luxa[1], Fedor Lipilin[1], Kseniia Mosina[1], Vojtech Kundrat[4], Jakub Zalesak[5], Jana Vejpravova[6], Martin Zacek[6], Christoph Gadermaier[2], José J. Baldoví[3], Zdeněk Sofer[1§]

[1]Department of Inorganic Chemistry, University of Chemistry and Technology Prague, Technicka 5, 166 28 Prague 6, Czech Republic
[2]Dipartimento di Fisica, Politecnico di Milano Piazza Leonardo da Vinci 32, 20133 Milano, Italy
[3]Instituto de Ciencia Molecular (ICMol), Universitat de Valencia, Paterna 46980, Spain
[4]Department of Chemistry, Faculty of Science, Masaryk University, Kamenice 5, Brno 62500, Czechia
[5]Chemistry and Physics of Materials, University of Salzburg, Jakob-Haringer-Strasse 2A, 5020, Salzburg, Austria
[6]Department of condensed matter physics, Faculty of mathematics and Physics, Charles University, Ke Karlovu 5, 121 16, Prague 2, Czech Republic

\* These authors contributed equally to this work.
§ Correspondence should be addressed to: Zdenek Sofer (soferz@vscht.cz)

# 1. Strain dependence of Bandgap

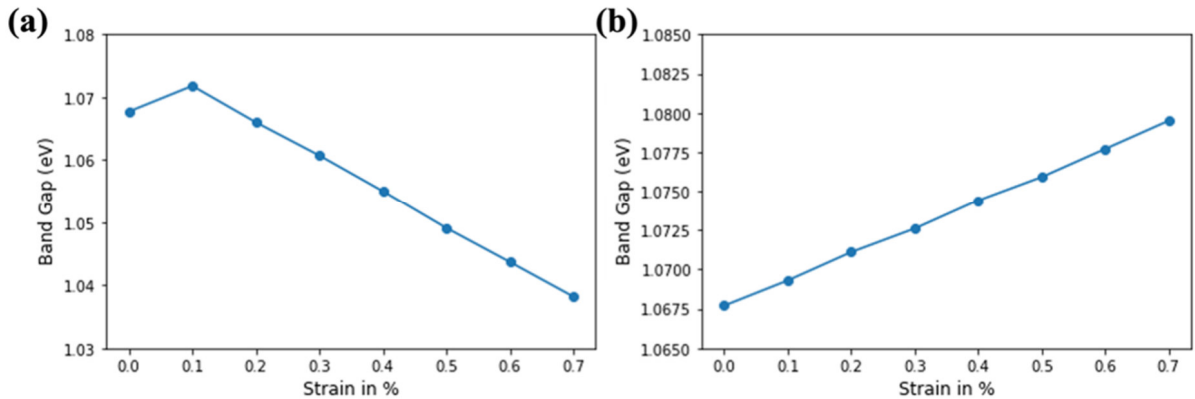

**Figure S1**. Theoretical prediction of change in band gap with tensile strain along (a) *a* (b) *b* axis of USe$_3$.

## 2. XRD surface morphology

The peaks observed in the X-ray diffraction (XRD) pattern taken on a single platelet aligned in the *ab*-plane match well with the out-of-plane 00L peaks for USe$_3$ from the ICDD database (Figure S2)[1]. For a full structural refinement USe$_3$ powder would have been required which we avoided for safety reasons and due to the sensitivity of USe$_3$ to oxygen and moisture. From the observed peaks we can only refine parameter $c = 10.455$ Å which is close to the literature value of 10.469 Å. The additional low-intensity peak at $2\theta = 42.36°$ does not match USe$_3$. We assign it to elemental selenium, consistent with metastable β-cubic Se observed in thin films and nanoscale precipitates.[2–4] The assignment is supported by (i) the ICDD match at this $2\theta$ (Figure S2), and (ii) SEM/EDS evidence of Se-rich nanospheres on the crystal surface often appearing along edges or grain boundaries (Figure S3 a-c and Figure 4a).

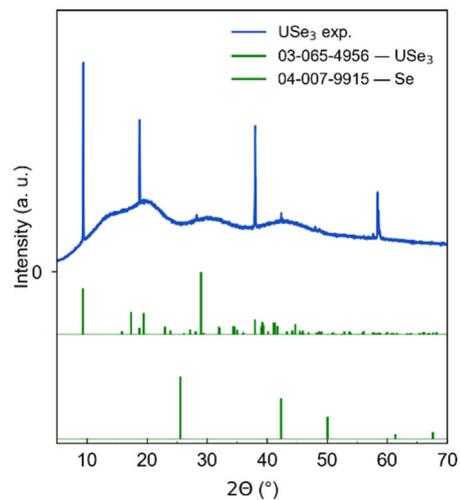

**Figure S2**. Raw XRD spectrum of USe$_3$ with literature ICDD cards[1] for USe$_3$ and cubic Se phase.

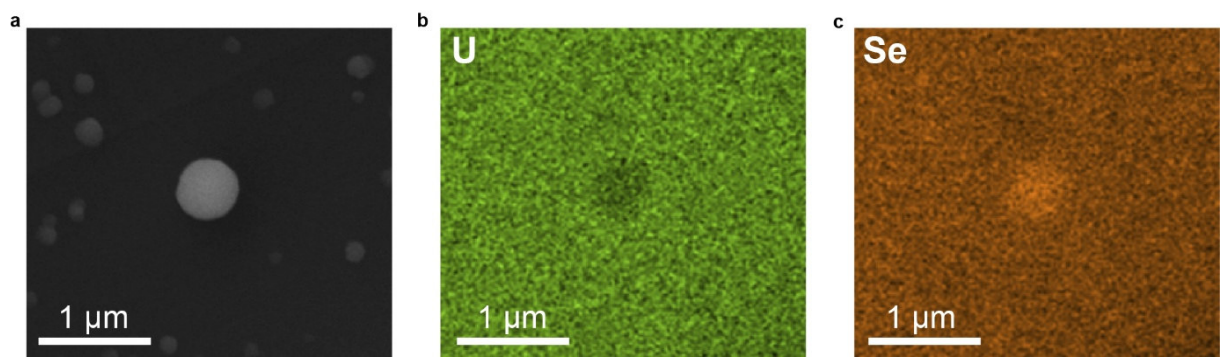

**Figure S3**. Energy dispersive X-ray spectroscopy (EDS) analysis of nanoscopic selenium spheres on the crystal surface. (a) SEM micrograph of sphere. (b) Uranium EDS map shows lack of U in the area of the sphere. (c) Selenium EDS map shows excess of Se in the area of the sphere.

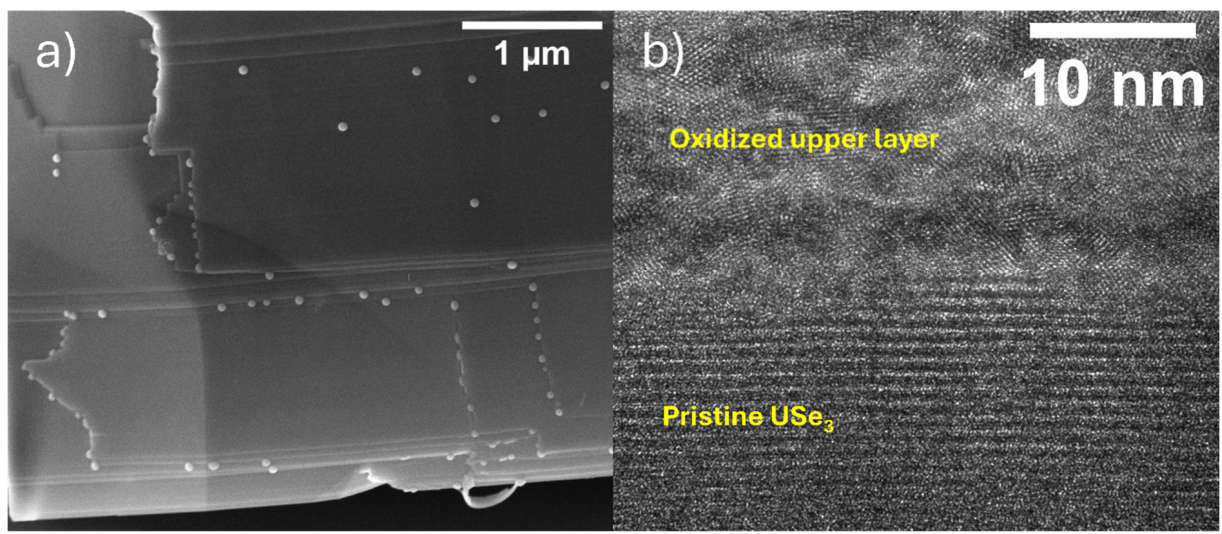

**Figure S4**. a) magnified SEM analysis of the surface of the USe$_3$ crystal showing a strap or band-like structure. Nanoscopic ball-like objects consist of elemental selenium. b) TEM analysis of the interface in the FIB cut lamella in Fig. 3d between the oxidized surface layer on the crystal and the pristine USe$_3$ phase.

## 3. EDS

EDS taken in a homogeneous region of a single crystal reveals a uniform distribution of uranium and selenium. The integrated composition across the area is U: 48.8 wt.% (27.8 at.%) and Se: 42.0 wt% (72.2 at.%) corresponding to a stoichiometry of U$_{1.16}$Se$_3$. The presence of oxygen or iodine (was used as transport agent during synthesis) is below 1%.

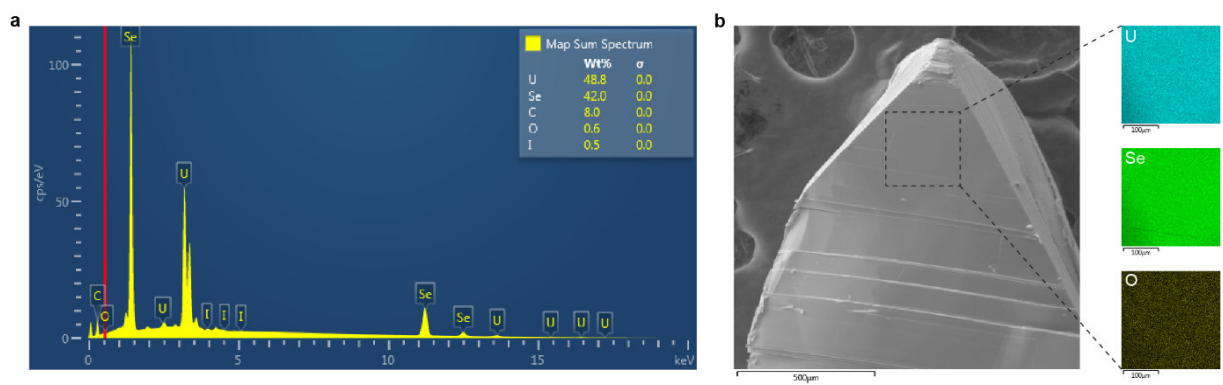

**Figure S5**. (a) EDS spectrum. (b) SEM micrograph of analyzed single crystal with dashed rectangle indicating the analyzed area. EDS maps show uniform distribution of U and Se and negligible presence of O.

## 4. Magnetic Measurements

A complication when studying USe$_3$ is its chemical instability and tendency to oxidize or undergo surface degradation. As a result, magnetic anomalies from previous reports, especially from powder samples, may reflect extrinsic effects from sample deterioration rather than intrinsic material properties. To address this, we prepared single-crystalline USe$_3$ samples in an argon atmosphere and minimized their air exposure during mounting and measurement. This approach improves structural integrity and yields cleaner magnetic signatures, enabling a more reliable assessment of the material's inherent anisotropy.

Previous magnetic investigations of monoclinic uranium trichalcogenides, including UTe$_3$, USe$_3$, and US$_3$, revealed that the inverse magnetic susceptibility does not follow the Curie–Weiss law, and instead exhibits pronounced anomalies: both UTe$_3$ and USe$_3$ show broad maxima in their magnetic susceptibility at 19.5 K and 50 K, respectively, while US$_3$ displays temperature-independent paramagnetism below approximately 15 K. The relatively large U–U separations (4.561 and 4.056 Å for USe$_3$) indicate that the 5$f$ electrons of the U$^{4+}$ ions are fairly localized, and that magnetic ordering likely arises from super exchange interactions mediated by the anions. It has been widely accepted that the low crystal symmetry characteristic of this family (monoclinic, space group P2$_1$/m, ZrSe$_3$-type structure) does not support conventional long-range magnetic ordering.[5–9]

In the main manuscript Figure 2e, we show the temperature dependence of magnetization, revealing significant anisotropy between the two in-plane directions. All curves feature a broad anomaly at 50 K, 44 K, and 40 K for the magnetic field applied along $a$, $b$, and $c$ directions, respectively. These broad maxima are usually ascribed to the transition of the paramagnetic to a long-range antiferromagnetically ordered state. The observed values are within the range reported in the literature.[6,7,9]

Furthermore, in Figure 2f, we show the temperature dependence of the inverse molar magnetic susceptibility. Here, data shows significant deviation from the Curie-Weiss law, which can be attributed to the strong crystal field acting on the electronic states of the U(IV) ion. [6,10]

We further show in figure S6 that the magnetization isotherms measured at 2 K are linear up to 14 T, and the magnetic moment per formula unit reaches about 0.15 – 0.30 B.M. at the maximum applied field of 14 T. This value is far below the expectation for a free U(IV) ion, corroborating the pronounced effect of the crystal field on the U(IV) electronic levels, observed in the susceptibility data. Such reduction is, however, overall common in this class of uranium materials.[6,10]

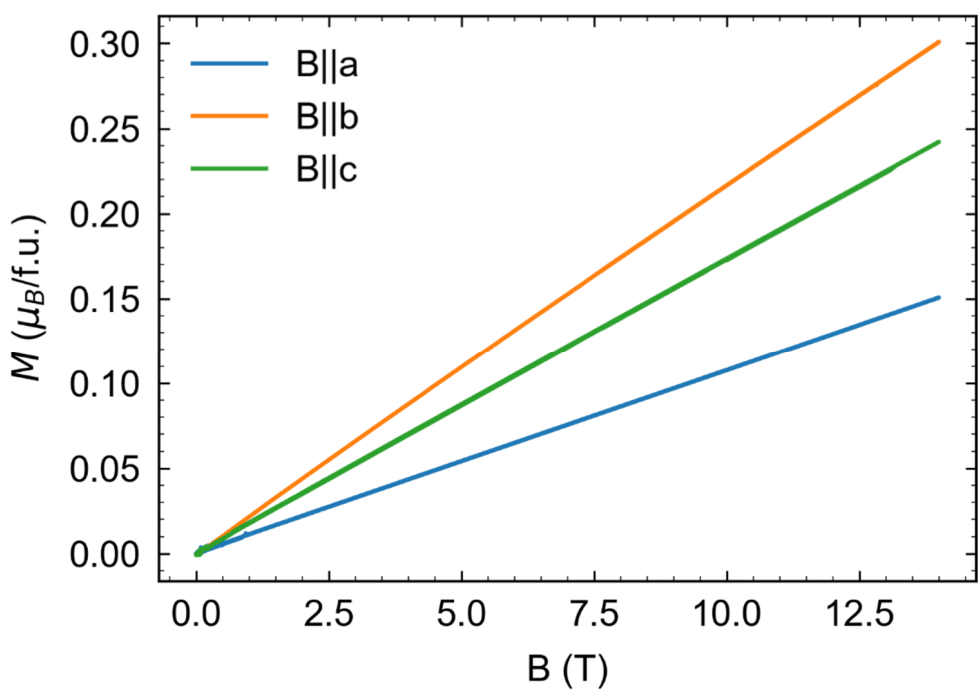

**Figure S6**. **a**. Magnetization isotherms of USe3 for the magnetic field applied along the three principal crystallographic axes at $T = 2$ K.

Our study of single-crystalline USe$_3$ reveals clear signatures of anisotropic magnetic behavior and long-range antiferromagnetic ordering, marked by broad susceptibility maxima near 50 K. The strong deviation from Curie–Weiss behavior and low magnetic moment values highlight the critical roles of crystal field effects and 5*f* electron localization. By minimizing sample degradation, we provide intrinsic magnetic data that confirm and refine earlier findings. These results position USe$_3$ as a representative example of low symmetry 5*f* electron systems, where anisotropy and electronic correlations govern the magnetic ground state.

## 5. Polarized Raman

Figure S7 shows the Raman spectrum measured with incident polarization along the *a*-axis (E || *a*). All resolvable Raman modes are fitted with a single Gaussian profile; $A_g^{(8)}$ at 283.3 cm$^{-1}$ is fitted with two deconvoluted Gaussian profiles to determine the position of the amorphous selenium mode presenting as a low energy shoulder to the Ag mode. The peak positions are summarized in Table S1 together with the calculated modes. The positions of the experimental modes are overall in good agreement with the theoretical predictions except for $B_g^{(3)}$ where a mismatch of 23.8 cm$^{-1}$ was observed.

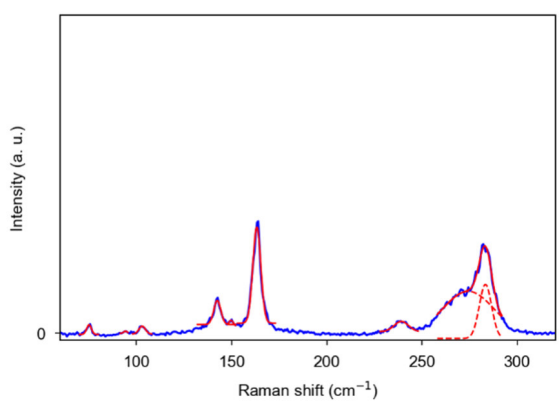

**Figure S7**. Raman spectrum with peak positions determined by peak fitting.

**Table S1**. Experimental and theoretical Raman modes.

|  | $A_g^{(1)}$ | $B_g^{(1)}$ | $A_g^{(2)}$ | $A_g^{(3)}$ | $B_g^{(2)}$ | $B_g^{(3)}$ | $B_g^{(4)}$ | $A_g^{(4)}$ | $A_g^{(5)}$ | $A_g^{(6)}$ | $A_g^{(7)}$ | $A_g^{(8)}$ | Se (cr.) | Se (am.) |
|---|---|---|---|---|---|---|---|---|---|---|---|---|---|---|
| *exp.* (cm$^{-1}$) | - | - | - | 75.0 | 94.2 | 103.0 | - | 142.4 | 149.7 | 163.2 | - | 283.3 | 239.1 | 274.3 |
| *theo.* (cm$^{-1}$) | 35.7 | 45.3 | 52 | 74.6 | 90.7 | 127.2 | 140.8 | 141.8 | 147.8 | 160.8 | 179.7 | 281.2 | - | - |

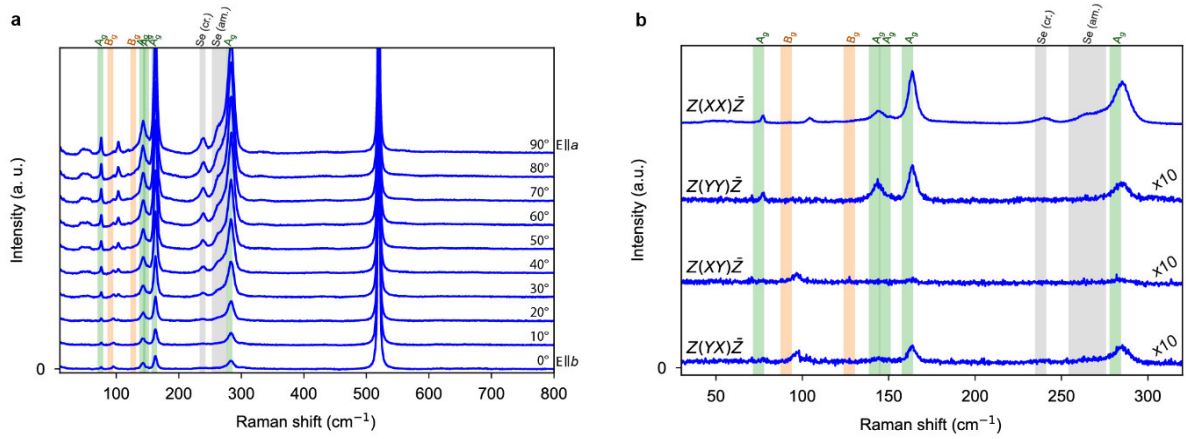

**Figure S8**. **a**. Full Raman spectrum with angle dependent excitation. **b**. Fully polarized Raman spectra of the low energy region.

Figure S8a shows the full Raman window of angle dependent Raman spectra demonstrating the absence of additional modes beyond 300 cm$^{-1}$ and the repeatability of the measurement in Figure 4d. Here we also observe more clearly that $B_g^{(2)}$ at 94.2 cm$^{-1}$ does not appear to be affected by the incident light polarization angle. To further investigate this anomaly we carried out fully polarization-resolved measurements (Figure S8b) by adding a fixed analyzer into the backscattered beam path set either parallel or perpendicular to the incident polarization we probe the four polarization configurations $Z(XX)\bar{Z}$, $Z(YY)\bar{Z}$, $Z(XY)\bar{Z}$, $Z(YX)\bar{Z}$ according to Porto notation. The $B_g^{(2)}$ still shows no dependence on the incident polarization but is completely suppressed in the parallel configuration and only visible in both cross-polarized configurations with equal intensity.

## 6. Thickness dependence of phonon modes

Raman spectra were taken on various flakes of USe$_3$ on Si/SiO$_2$ with a thickness ranging between 3 nm and 50 nm measured using atomic force microscopy (AFM). The peak positions of the 4 most intense modes ($A_g^{(3)}$, $A_g^{(5)}$, $A_g^{(6)}$, and $A_g^{(8)}$) are plotted against their thickness in Figure S9. No clear thickness dependence is observed.

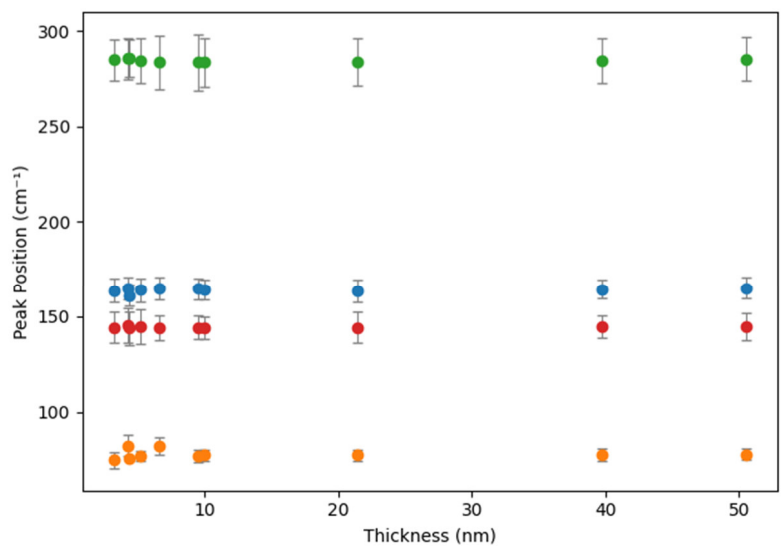

**Figure S9**. Peak positions of 4 $A_g$ modes vs. thickness determined via AFM. No clear thickness dependence for these modes visible in the observed range.

## 7. Raman Tensors

We report Density Functional Perturbation Theory (DFPT) Raman tensors in a cartesian frame with $x \parallel a$, $y \parallel b$, and $z \perp ab$ (laser direction). Tensor units are $\text{Å}^2/\sqrt{\text{amu}}$.

$$\mathbf{R}(A_g^{(1)}) = \begin{pmatrix} -40.764 & -0.335 & -1.276 \\ -0.328 & 0.060 & -0.410 \\ 1.211 & -0.410 & 0.134 \end{pmatrix} \quad \mathbf{R}(B_g^{(1)}) = \begin{pmatrix} 29.298 & 0.039 & -1.104 \\ -0.024 & 0.140 & -0.210 \\ -0.634 & -0.211 & 1.450 \end{pmatrix}$$

$$\mathbf{R}(A_g^{(2)}) = \begin{pmatrix} 3.434 & 0.000 & 1.070 \\ 0.000 & -0.388 & 0.000 \\ -0.400 & 0.000 & 0.643 \end{pmatrix} \quad \mathbf{R}(A_g^{(3)}) = \begin{pmatrix} 19.327 & -0.000 & 0.318 \\ 0.000 & 1.238 & 0.000 \\ -0.074 & 0.000 & 1.455 \end{pmatrix}$$

$$\mathbf{R}(B_g^{(2)}) = \begin{pmatrix} 0.224 & 0.044 & -1.698 \\ 0.046 & -0.124 & 0.037 \\ -1.346 & 0.034 & -0.180 \end{pmatrix} \quad \mathbf{R}(B_g^{(3)}) = \begin{pmatrix} 6.743 & -0.000 & -1.376 \\ 0.000 & -0.153 & 0.000 \\ -1.429 & 0.000 & 1.457 \end{pmatrix}$$

$$\mathbf{R}(B_g^{(4)}) = \begin{pmatrix} 17.218 & 0.292 & 1.795 \\ 0.296 & 0.020 & 0.346 \\ -1.246 & 0.346 & -0.112 \end{pmatrix} \quad \mathbf{R}(A_g^{(4)}) = \begin{pmatrix} 46.193 & -0.048 & 2.301 \\ 0.029 & -0.348 & 0.259 \\ 1.934 & 0.260 & -1.497 \end{pmatrix}$$

$$\mathbf{R}(A_g^{(5)}) = \begin{pmatrix} -4.653 & 0.000 & -0.590 \\ 0.000 & 0.519 & -0.000 \\ 0.286 & 0.000 & -0.971 \end{pmatrix} \quad \mathbf{R}(A_g^{(6)}) = \begin{pmatrix} -7.184 & -0.000 & -1.375 \\ 0.000 & -1.079 & 0.000 \\ -0.761 & 0.000 & -1.365 \end{pmatrix}$$

$$\mathbf{R}(A_g^{(7)}) = \begin{pmatrix} -21.085 & -0.038 & -0.764 \\ -0.041 & 0.092 & -0.030 \\ 1.359 & -0.030 & 0.248 \end{pmatrix} \quad \mathbf{R}(A_g^{(8)}) = \begin{pmatrix} 43.313 & -0.000 & -1.712 \\ 0.000 & 0.057 & 0.000 \\ -1.876 & 0.000 & -1.330 \end{pmatrix}$$

For incident and analyzed unit polarization vectors $\mathbf{e}_{in}$ and $\mathbf{e}_{out}$ the Raman intensity of a single Stokes mode is

$$I(\mathbf{e}_{in}, \mathbf{e}_{out}) \propto |\mathbf{e}_{out}^T \mathbf{R} \, \mathbf{e}_{in}|^2.$$

Most of our computed modes are dominated by diagonal elements, specifically the $\mathbf{R}_{xx}$ element, which explains the strong response for $E \parallel a$ in parallel geometry and near extinction for $E \parallel b$. In contrast, the $B_g^{(2)}$ mode stands out for its comparatively large off-diagonal elements and comparable magnitudes of $\mathbf{R}_{xx}$ and $\mathbf{R}_{yy}$ which explains the observed insensitivity to incident light polarization and high intensity in cross-polarized geometry.

## 8. Further Strain Measurements

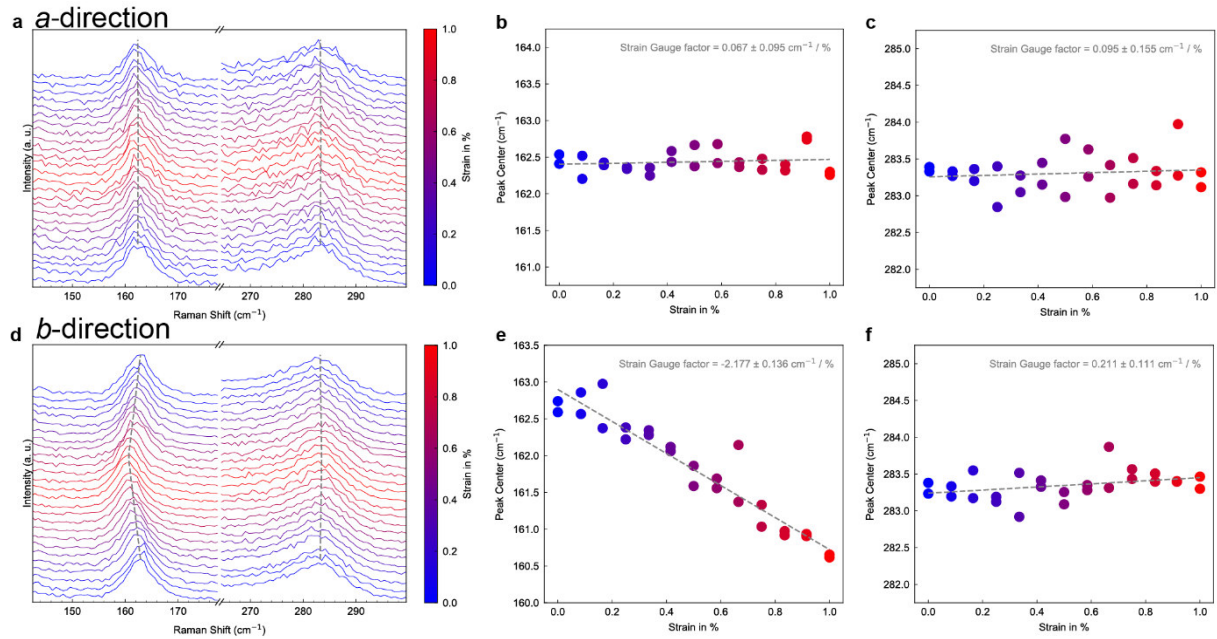

**Figure S10**. **a**. Partial Raman spectra showing $A_g^{(6)}$ and $A_g^{(8)}$ modes of exfoliated USe$_3$ on PC for values of strain between 0% and 1.0% along the *a*-direction. **b.** Peak center vs. strain plot of $A_g^{(6)}$ mode indicates no shift. **c.** Peak center vs. strain plot of $A_g^{(8)}$ mode indicates no shift. **d**. Partial Raman spectra showing $A_g^{(6)}$ and $A_g^{(8)}$ modes of exfoliated USe$_3$ on PC for values of strain between 0% and 1.0% along the *b*-direction. Here the $A_g^{(6)}$ mode shows a clear shift. **e.** Peak center vs. strain plot of $A_g^{(6)}$ mode indicates a clear shift. **f.** Peak center vs. strain plot of $A_g^{(8)}$ mode indicates no shift.

## 9. *a*-direction strain causing delamination

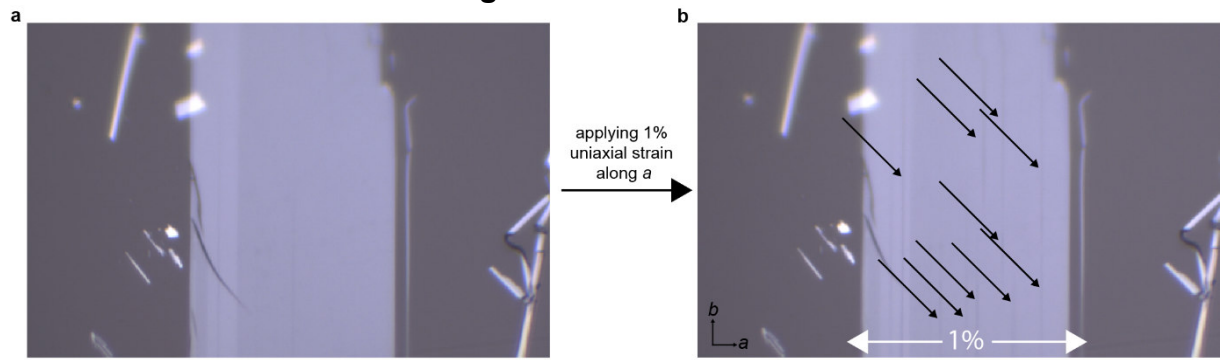

**Figure S11**. Delamination under uniaxial strain along *a*-direction. a. unstrained sample. b. Same sample after 1% uniaxial strain along the *a*-direction was applied. Black arrows indicate tears which formed inside the sheet.

## 10. Strain Theory

We estimated the effect of tensile strain along the *a* and *b* axis. For the strain along the *b* axis, we can clearly observe a red shift of an out-of-plane vibrational $A_g^{(6)}$ mode (160.8 cm$^{-1}$), consistent with the experiment (table S2). On the other hand, the in-plane vibrational $A_g^{(8)}$ mode (281.2 cm$^{-1}$) does not show a significant shift with tensile strain along the b-axis, which is also in agreement with the experiment (table S2). Contrary to this, with the tensile strain along the *a*-axis, the $A_g^{(6)}$ mode exhibits a small redshift, while the $A_g^{(8)}$ mode shows a significant redshift. This clearly indicates the anisotropic nature of the materials.

**Table S2**: Phonon energies of simulated Raman modes of USe$_3$ with strain (in %) along the *b* crystallographic axis.

| Symmetry | Strain | | | | | | | |
|---|---|---|---|---|---|---|---|---|
| | 0.0 | 0.1 | 0.2 | 0.3 | 0.4 | 0.5 | 0.6 | 0.7 |
| Ag | 35.7 | 35.9 | 37.6 | 35.9 | 37.2 | 33.2 | 35.7 | 38.3 |
| Bg | 45.3 | 45.3 | 45.0 | 45.0 | 45.0 | 44.9 | 44.8 | 44.7 |
| Ag | 52.0 | 52.5 | 52.0 | 52.5 | 52.3 | 52.3 | 53.9 | 53.1 |
| Ag | 74.6 | 74.6 | 74.3 | 74.2 | 74.1 | 74.1 | 73.9 | 73.8 |
| Bg | 90.7 | 90.3 | 89.8 | 89.4 | 89.1 | 88.6 | 88.3 | 87.7 |
| Bg | 127.2 | 127.1 | 126.8 | 126.6 | 126.4 | 126.2 | 126.0 | 125.6 |
| Bg | 140.8 | 140.8 | 140.6 | 140.0 | 140.1 | 139.6 | 139.2 | 138.8 |
| Ag | 141.8 | 141.5 | 142.0 | 141.8 | 141.6 | 141.3 | 141.0 | 140.7 |
| Ag | 147.8 | 147.6 | 147.2 | 146.5 | 146.3 | 145.7 | 145.5 | 145.2 |
| Ag | 160.8 | 160.5 | 160.0 | 159.6 | 159.3 | 158.9 | 158.6 | 158.1 |
| Ag | 179.7 | 179.4 | 178.8 | 178.4 | 178.0 | 177.6 | 177.3 | 176.8 |
| Ag | 281.2 | 281.1 | 281.3 | 281.2 | 281.1 | 281.0 | 280.8 | 281.0 |

**Table S3**: Phonon energies of simulated Raman modes of USe$_3$ with strain (in %) along *a* crystallographic axis.

| Symmetry | Strain | | | | | | | |
|---|---|---|---|---|---|---|---|---|
| | 0.0 | 0.1 | 0.2 | 0.3 | 0.4 | 0.5 | 0.6 | 0.7 |
| Ag | 35.7 | 37.2 | 36.7 | 38.3 | 39.6 | 39.1 | 38.0 | 38.6 |
| Bg | 45.3 | 45.1 | 45.3 | 45.2 | 45.2 | 45.2 | 45.2 | 45.1 |
| Ag | 52.0 | 51.7 | 52.9 | 52.7 | 53.1 | 53.1 | 52.5 | 52.5 |
| Ag | 74.6 | 74.5 | 74.5 | 74.6 | 74.4 | 74.8 | 74.6 | 74.6 |
| Bg | 90.7 | 90.9 | 91.3 | 91.5 | 91.8 | 92.0 | 92.2 | 92.5 |
| Bg | 127.2 | 127.2 | 127.1 | 127.1 | 126.9 | 126.9 | 126.9 | 126.7 |
| Bg | 140.8 | 140.6 | 140.5 | 140.3 | 140.1 | 140.0 | 139.8 | 137.7 |
| Ag | 141.9 | 141.7 | 141.6 | 141.4 | 141.3 | 141.2 | 141.0 | 139.4 |
| Ag | 147.8 | 148.1 | 148.0 | 148.3 | 148.2 | 148.2 | 148.2 | 148.4 |
| Ag | 160.8 | 160.6 | 160.5 | 160.2 | 160.1 | 160.0 | 159.9 | 159.6 |
| Ag | 179.7 | 179.6 | 179.5 | 179.5 | 179.4 | 179.5 | 179.4 | 179.0 |
| Ag | 281.2 | 280.0 | 278.6 | 277.3 | 275.9 | 274.6 | 273.1 | 271.7 |

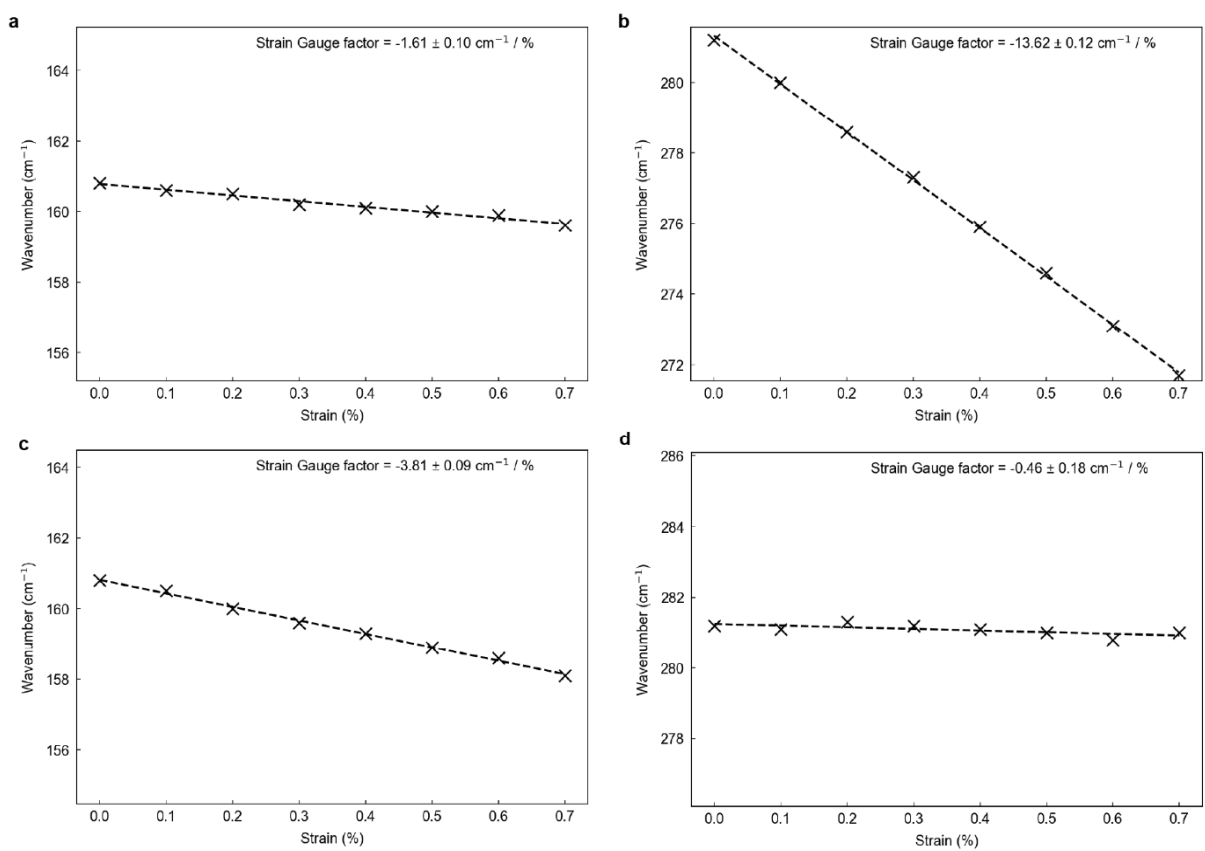

**Figure S12.** Calculated phonon mode shifts under uniaxial strain. **a**, **b** Strain applied along the crystallographic *a* direction. **c**, **d** strain applied along the *b* direction. Each for the modes at 160.8 cm$^{-1}$ and 281.2 cm$^{-1}$. Black crosses indicate calculated frequencies at each applied strain, and dashed black lines are linear fits. Annotated values represent the strain gauge factor with standard error of the slope.

## 11. Electrical Transport

**Table S4: Device Parameters**

|          | $W$ / μm | $L$ / μm | $t_{\text{ins.}}$ / nm | $t_{\text{ch}}$ / nm |
|----------|----------|----------|------------------------|----------------------|
| Device *a* | 18.4     | 5.0      | 29.8                   | 32.2                 |
| Device *b* | 11.0     | 9.8      | 11.5                   | 18.2                 |

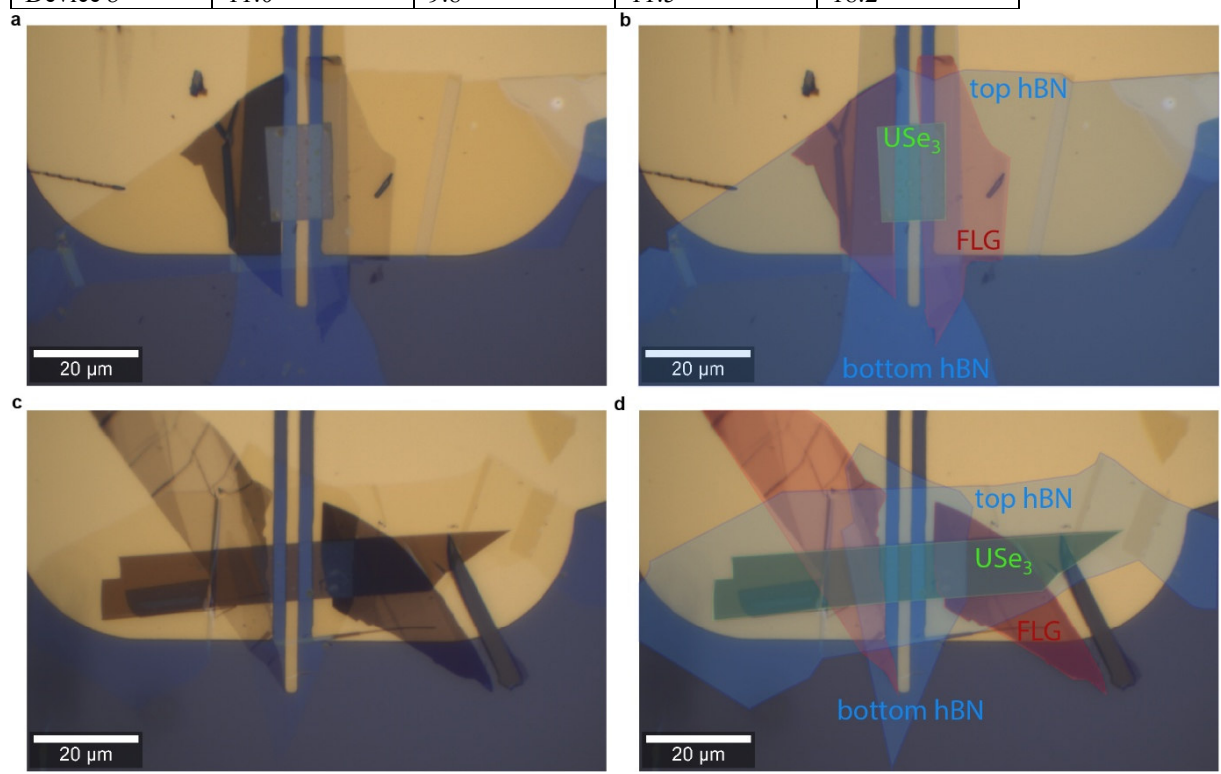

**Figure S13** Field Effect Transistor **(**FET) devices. a. High resolution optical micrograph of *a*-direction device. b. Annotated version to help identify the individual components: USe$_3$ channel (green), top and bottom hBN encapsulants (blue), few-layer graphene (FLG) contacts (red). c. High resolution optical micrograph of *b*-direction device. d. Annotated version to help identify the individual components: USe$_3$ channel (green), top and bottom hBN encapsulants (blue), few-layer graphene contacts (red).

## 12. Electrical Transport Setup Theory

The transport computation setup is shown in Figure S14. The source and drain of the system are semi-infinite metallic leads that are coupled to a central USe$_3$ area. The carrier concentration is modulated via a gate electrode. Electrical conductivity exhibits substantial anisotropic behavior when the transport computations are carried out along the *a* and *b* crystallographic directions.

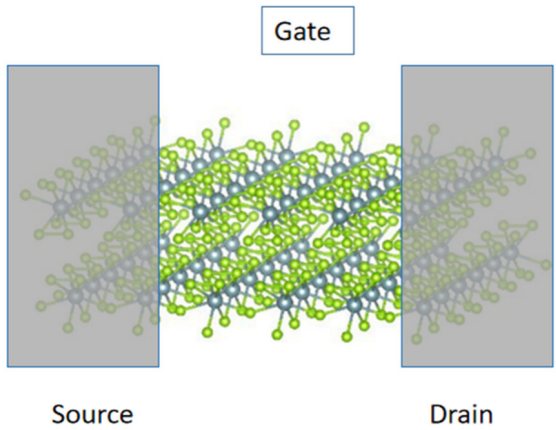

**Figure S14**: Schematic of the transport system showing the source, drain, and gate electrodes,
for USe$_3$ material.

The electrical conductivity (σ) along the *a* and *b* directions is estimated using:

$$\sigma = (I_{DS} / V_{DS}) \times \left(\frac{L}{A}\right)$$

where $I_{DS}$ is the computed current, $V_{DS}$ is the applied bias voltage. *L* is the transport channel length, and *A* is the cross-sectional area.

Theoretical Transport Properties and Anisotropy The computed mobility values at -6V gate bias are: $\mu_a = 4.85 \cdot 10^{-4}$ cm²/Vs, $\mu_b = 1.62 \cdot 10^{-4}$ cm²/Vs, indicating that transport is significantly more efficient along the *a* direction than the *b* direction.

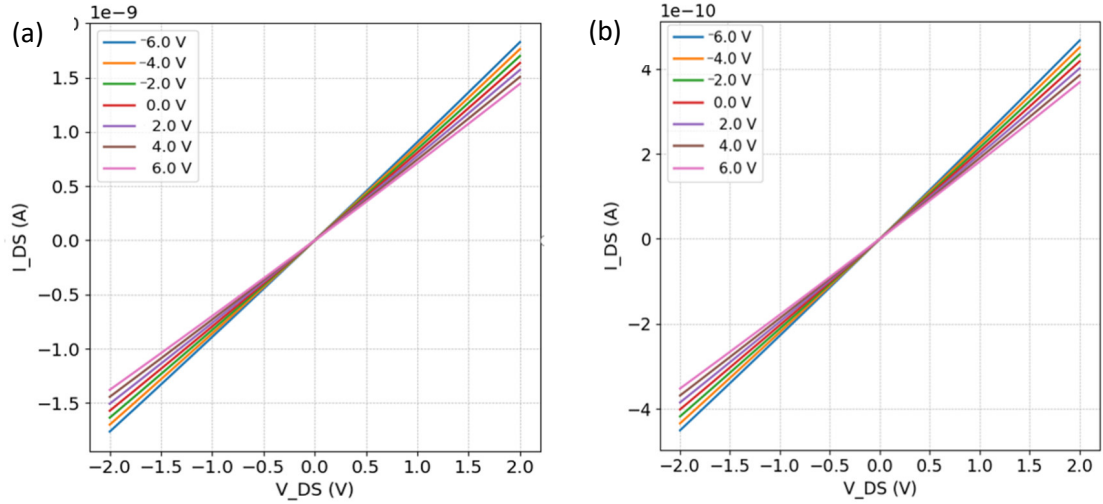

**Figure S15**: Theoretical I-V Characteristics curves for devices oriented along *a*-direction (a) and *b*-direction for various gate voltages.

Electrostatic tuning of carrier transport is seen by the shift in curves with different gate voltages. This change is explained by the applied gate voltage modulating the energy levels in the valence and conduction bands, which alters the charge carrier density and, consequently, the current. The validity of the NEGF approach in conjunction with DFT-based Hamiltonian extraction is confirmed by the concordance between theoretical and experimental results.

The I–V curves show Ohmic behavior, staying linear up to ±2 V and indicating minimal Schottky barriers. Increasing |VG| raises the current amplitude, matching hole-dominated conduction in USe$_3$. At the same |VG|, the channel along *a* has a steeper slope than along *b*, yielding a slightly higher conductivity and mobility. Compared to other semiconductors the conductivity is low because uranium interactions localize electrons, imperfections cause carrier scattering, and low Fermi-level Density of States (DOS) limits mobility.